\documentclass[oldversion]{aa}
\usepackage{txfonts}
\usepackage{graphicx}
\ignorespaces

\def\TMB {$T_{\rm mb} $ }
\def\TKIN {$T_{\rm kin} $ }

\def\H0 {$H_{\rm o}$}
\def\mic {$\mu\hbox{m} $ }

\def\solmass {\hbox{$\rm M_{\odot}  $} }

\def\solum {\hbox{$\rm L_{\odot}  $} }

\def\numd {\hbox{$n\,({\rm H}_2$) } }                   

\def\kms {\hbox{${\rm km\,s}^{-1} $} }

\def\Kkms {\hbox{${\rm K\,km\,s}^{-1} $}}
\def\percc {$\hbox{{\rm cm}}^{-3} $ }    
\def\cmsq  {$\hbox{{\rm cm}}^{-2} $ }    
\def\arcsec {\hbox{$^{\prime\prime} $} }

\def\NH3 {\hbox{${\rm NH}_{3}$}}                  
\def\TCO {\hbox{${\rm ^{12}CO}  $} }               
\def\THCO {\hbox{$^{13}{\rm CO}  $} }              
\def\CH3C2H {\hbox{${\rm CH}_3{\rm C}_2{\rm H}$}} 
\def\CH3OD {\hbox{${\rm CH}_3{\rm OD}$}}          
\def\HC15N {\hbox{${\rm HC}^{15}{\rm N}$}}        
\def\HN13C {\hbox{${\rm HN^{13}C}$}}              
\def\SO2 {\hbox{${\rm SO_{2}}$}}                  
\def\H2S {\hbox{${\rm H_{2}S}$}}                  
\begin{document}
    \title{The interstellar medium of the Antennae Galaxies} 
 
   \author{A.\ Schulz\inst{1,3} 
\and C.\ Henkel\inst{2}
\and D.\ Muders\inst{2}
\and R.Q.\ Mao\inst{4}
\and M.\ R\"ollig\inst{3,5}
\and R.\ Mauersberger\inst{6}
           }
\offprints{A.\ Schulz, andreas.schulz@uni-koeln.de}

        \institute{Institut f\"ur Physik und ihre Didaktik, Universit\"at zu 
           K\"oln, Gronewaldstr. 2, D-50931 K\"oln, Germany 
        \and
           Max-Planck-Institut f\"ur Radioastronomie, Auf dem H\"ugel 69, 
           D-53121 Bonn, Germany 
        \and
           Argelander-Institut f\"ur Astronomie, Universit\"at 
           Bonn, Auf dem H\"ugel 71, D-53121 Bonn, Germany 
        \and
           Purple Mountain Observatory, Chinese Academy of Sciences, 
           210008 Nanjing, P.R. China
        \and
           I. Physikalisches Institut, Universit\"at zu K\"oln, 
           Universit\"atsstr. 17, D-50937 K\"oln, Germany
        \and
           Instituto de Radioastronom{\'i}a Milim\'etrica (IRAM), 
           Avenida Divina Pastora 7, Local 20, E-18012 Granada, Spain
                        }

   \date{Received: June 2006 ; accepted: October 24, 2006 }

\abstract{
To study the properties of the interstellar medium in the prototypical merging system 
of the Antennae galaxies (NGC\,4038 and NGC\,4039), we have obtained \TCO (1--0), 
(2--1) and (3--2) line maps, as well as a map of the 870$\mu$m continuum emission. 
Our results are analysed in conjunction with data from X-ray to radio wavelengths. 
In order to distinguish between exact coincidence and merely close correspondence of emission 
features, we compare the morphological structure of the different emission  
components at the highest available angular resolution. 
To constrain the physical state of the molecular gas, we apply models of photon dominated 
regions (PDRs) that allow us to fit CO and 
[C{\sc ii}] data, as well as other indicators of widespread PDRs in the Antennae system, 
particularly within the super giant molecular cloud (SGMC) complexes of the interaction region 
(IAR) between the two galaxies. The modeled clouds 
have cores with moderately high gas densities up to 4 10$^4$ \percc and rather low kinetic 
temperatures ($\leq$25K). At present, all these clouds, including those near the galactic nuclei, 
show no signs of intense starburst activity. Thermal radio or mid-infrared emission are all 
observed to peak slightly offset from the molecular peaks. 
The total molecular gas mass of the Antennae system adds up to $\sim$10$^{10}$ M$_{\odot}$.
In the vicinity of each galactic nucleus, the molecular gas mass, 1 -- 2 10$^9$ M$_{\odot}$, 
exceeds that of the Galactic 
centre region by a factor of almost 100. Furthermore, the gas does not seem to deviate much from the 
$N_{\rm H_2}$/$I_{\rm CO}$ ratio typical of the disk of our Galaxy rather than our Galactic centre. 
Alternative heating mechanisms with respect to PDR heating are discussed. 
}


\maketitle

\section{Introduction}

Although contributing only a fraction of the total mass, interstellar matter  
plays a profound role in mergers of spiral galaxies. Their molecular gas 
fuels an intense starburst and enormous far-infrared (FIR) luminosities are reached 
(Sanders \& Mirabel 1996).  
Major mergers between gas-rich galaxies of comparable mass trigger 
the most luminous starbursts known. Galaxy interactions occurred frequently 
in the early universe. Hence, studying such interactions 
will contribute to our understanding of early galaxy evolution as well. 

The Antennae galaxies (NGC\,4038/39, Arp 244) 
are the nearest example of such a strongly interacting pair of galaxies. 
For the distance of the system (see Stanford et al. 1990; Wilson et al. 2000;  
Gao et al. 2001) we adopt 20 Mpc (1$''$ covers $\sim$100 pc). 
Believed to be a symmetric encounter between two normal Sc type spirals, 
the Antennae have been well-studied at most wavelengths and are the 
subject of extended numerical simulations. 
The Antennae system is thought to be a merger in its early or possibly 
intermediate stages.
(1) Optical observations reveal that the 
highly active star-forming regions are placed in an area between the two 
nuclei where a population of hundreds of supermassive stellar clusters 
is found. The majority of these  have ages of $\sim$4 10$^8$ yr or 
less (Fritze-von-Alvensleben 1999) and the brightest of them are even 
younger than $\sim$10$^7$ yr, with a part of their starlight being 
heavily reddened hinting at large amounts of dust. Infrared Space Observatory 
(ISO) observations revealed luminous regions of massive star formation that are 
invisible at optical wavelengths (Mirabel et al. 1998).
Their stellar content strongly surpasses that of open stellar clusters 
(Whitmore \& Schweizer 1995; Zhang et al. 2001).  
(2) In the same region one finds the maxima of FIR emission (Evans et al. 1997; 
Bushouse et al. 1998) and submm emission (Haas et al. 2000) that are believed 
to originate in warm dust. (3) This is accompanied by strong peaks of
radio emission observed at high angular resolution (Hummel \& van der Hulst 1986; 
Neff \& Ulvestad 2000) with the individual sources being  partly thermal and 
coinciding with optical/H${\alpha}$ knots.

Physical and kinematical properties of the molecular gas  
are important for our understanding of the triggering of intense star formation in 
galaxy mergers. Previous observations of CO emission in the  
Antennae (e.g. Wilson et al. 2000) reveal extremely massive concentrations of  
molecular gas ($\ge$10$^8$\,\solmass), a chaotic velocity field, and evidence 
of cloud-cloud collisions near the strongest mid-infrared peak. 

Gao et al. (2001) claim that, while starbursts are observed  
in some extended regions in the Antennae, the most intense phase of the starburst is 
only just beginning. They predict that the Antennae ($L_{\rm FIR}$ $\sim$ 10$^{11}$\solum 
currently) will undergo the much stronger ultraluminous 
($L_{\rm FIR}$ $\ga$10$^{12}$L$_{\odot}$) starburst phase.
Their arguments are based on the assumption that the star formation can be 
modeled using a Schmidt law, the consequences of which have been analysed 
for Antennae-like systems (Mihos \& Hernquist 1996).  Assessing these claims 
requires an accurate analysis of the interstellar gas in its various phases.

Furthermore, several investigations (e.g. Nikola et al. 1998; Fischer et al. 
1996) raise evidence of the widespread presence of  {\bf P}hoton 
{\bf D}ominated {\bf R}egions ({\bf PDR}s) in merging systems. This is not 
unexpected in view of a starburst, initiated a few 10 Myr ago, that must have 
produced large quantities of B stars radiating soft UV photons at energies 
below 13 eV. For the central 
parts of two rather extreme cases of nearby galaxies, namely the 
``quiescent'' galaxy IC\,342 and the ``starburster'' M\,82, we have shown that 
the thermal budget of interstellar molecular gas can be explained predominantly 
in terms of a PDR scenario (Schulz et al. 2001; Mao et al. 2000). The PDR scenario 
for the molecular gas in M\,82 is confirmed by investigations of Garcia-Burillo 
et al. (2002), Fuente et al. (2005), and Fuente et al (2006) on the basis of HCO, 
HCO$^+$, and CO$^+$ observations, respectively. PDRs are also found in the spiral 
arms of M\,83 and M\,51 (Kramer et al. 2005). 

After a brief kinematical analysis of our data (Sect. 3.2.), we carry out an extended
inspection of the morphological structure of tracers for the different physical components 
of interstellar matter to examine their relationship to the molecular clouds (Sect. 4). 
This will be succeeded by an excitation analysis of the gas contained in the 
molecular clouds (Sect. 5). It is our aim to investigate whether a PDR model is applicable to its 
present state followed by a discussion of other possible heating processes  for 
the bulk of the gas (Sect. 6).

\section{Observations}

The observations were performed during various  seasons from Feb. 1999 
to Apr. 2003 at the IRAM 30m-millimetre-radio-telescope (MRT) in Spain and at the 10m 
Heinrich-Hertz-Telescope (HHT) in Arizona, both
in the beam-switching mode using the chopping secondary mirror 
with a throw of 200$''$ (MRT) and 180$''$ (HHT) in azimuth 
at switching frequencies of 0.5 Hz (MRT) and 2 Hz (HHT). 
The integration time for a single measurement was typically 2 minutes; 
several positions including the central positions of NGC\,4038, 4039 and the 
{\bf I}nter{\bf A}ction {\bf R}egion ({\bf IAR}) were repeatedly measured to 
obtain total integration times of up to 30 minutes per position. 

A very good telescope pointing is essential for the complex cloud structure of the Antennae 
on arcsec scales; this was carefully checked every 30 to 40 minutes 
on  nearby continuum sources. IRAM MRT 
data with pointing offsets  of more than 1.5$''$  were ignored. 
The parallel alignment of the MRT receivers (at 3mm and 1mm)
was examined by observing planets, which also yielded the calibration checks. 
Hence, parallel observations of the CO\,(2--1) and (1--0) lines guarantee excellent 
relative pointing and therefore reliable line ratios. In addition, the image sidebands 
of the MRT receivers were suppressed by more than 10 dB using a Martin Pupplett interferometer 
filter. Furthermore, the telescope 
beams are very clean and free of prominent side lobes after improving the antenna, 
making it an ideal tool for the present study.

\begin{itemize}
\item 
{\bf CO\,(3--2)}: The 345 GHz SIS mixer receiver at the HHT had a  
receiver noise temperature of 150K (double side band). 
The main beam efficiency was 0.55. The main beam size 
was 22$''$ (full width of half maximum, FWHM). 
Total system noise temperatures (in the $T_{\rm mb}$ system) were typically 1500 -- 3000 K. 

\item
{\bf CO\,(2--1)}: Using the MRT  standard facility receivers at 230/220 GHz, 
calibration measurements of the improved telescope antenna 
yielded forward and main beam efficiencies of 0.91 and 0.52, respectively; 
the main beam size was 10.7$''$. Total system noise temperatures were 
about 700 -- 800 K. 

\item
{\bf CO\,(1--0)}: Observations with the 115/110 GHz IRAM standard facility receivers 
were performed in parallel with the CO\,(2--1) observations: forward and 
main beam efficiencies were 0.95 and 0.74, resp., beamwidths were 22$''$; the total system noise 
temperature was about 300 K.  

\item
{\bf 870$\mu$m continuum}: The HHT 19-channel bolometer array was used in its standard 
observing mode (see Mao et al. 2002 for details); the beamwidth was 22$''$. 

\end{itemize}
\smallskip

The standard facility IRAM and HHT backends were used with a typically 1 MHz 
spectral resolution. The final spectra were smoothed to channel spacings of  $\ge$6 kms$^{-1}$. 
Linear baselines were removed from the spectra. 

\smallskip

The intensities of the spectra are given in units of main beam brightness temperature ($T_{\rm mb}$). 
To derive line ratios useful for a well-constrained excitation analysis, absolute calibration 
of the spectral intensities had to be better than $\pm$10\%, which was achieved by observations 
of planets.

\section{Results}

\subsection{Spectra and maps}


Figure\,1 depicts the \TCO  spectra  at the IAR, and Figure\,2 
shows the map of integrated intensity of the CO\,(2--1) line with 10$''$ 
spacing, covering the entire target including the positions of both galactic centres 
and the IAR.  Figure\,3 presents CO\,(2--1) channel maps, and 
Figure\,4 displays the corresponding spectra of the inner part of the map allowing us to 
identify groups of cloud complexes at different velocities. Since the  
CO\,(2--1) beam covers an area with a linear size of 1 kpc, 
it is not possible to resolve individual  GMCs like e.g. SgrB2 in our own Galaxy.

The intensity distribution shows identical features within the noise for all three 
mapped CO lines. Most of the CO emission originates from the IAR, which is 
more extended than the nuclear regions.



\smallskip

   \begin{figure}  
   \centering
   \resizebox{8.5cm}{!}{\rotatebox[origin=br]{0}{\includegraphics{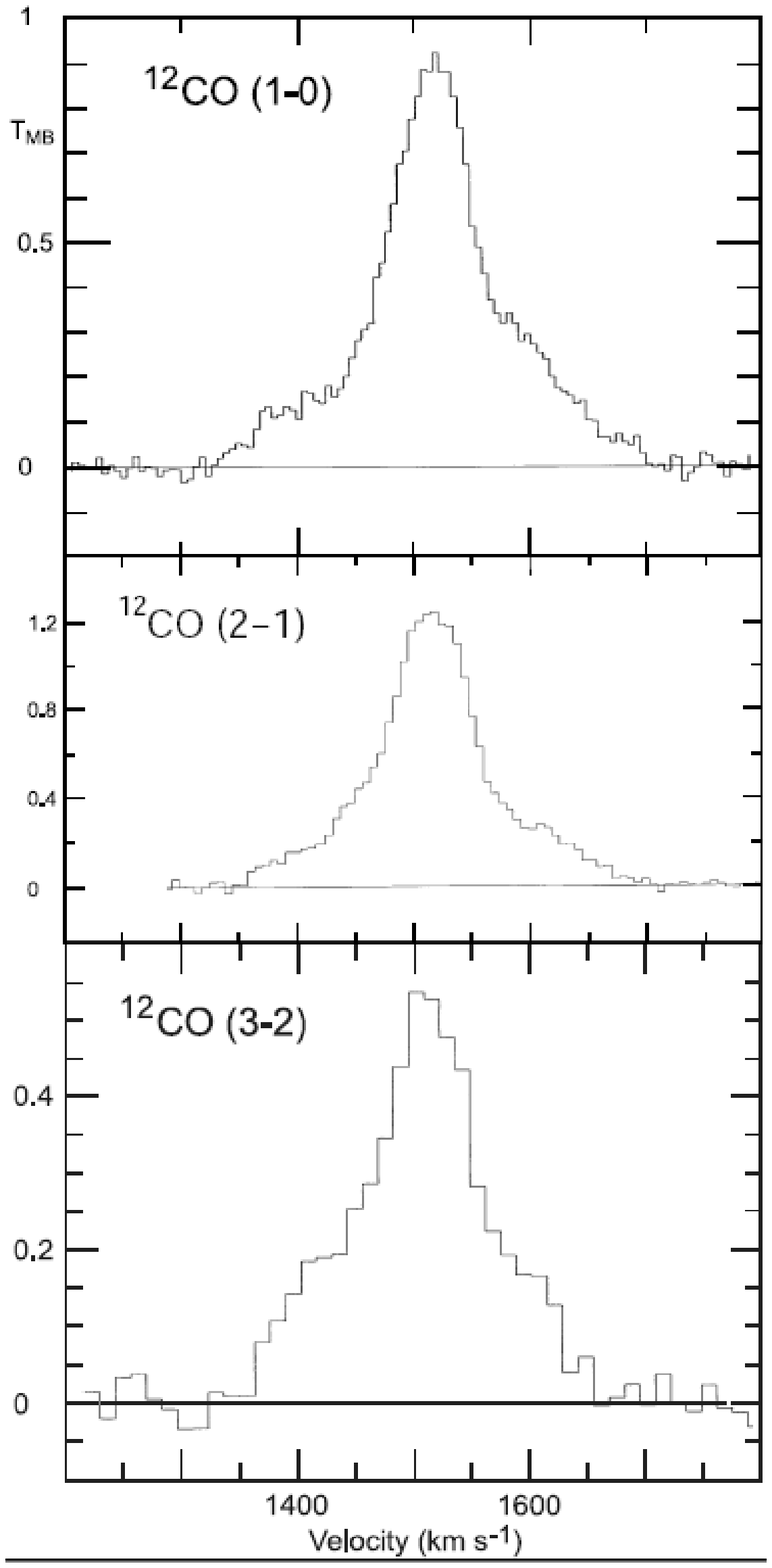}}}
   \caption{Observed spectra of \TCO\,(1--0), (2--1), and (3--2)   
   (channel spacing 6 and 13 kms$^{-1}$, respectively)  at the central position of the IAR 
  ((0,-10)-position of 
    Fig. 2.) 
    The intensity scale is in units of main beam brightness temperature, $T_{\rm mb}$.   }
   \label{fone}
\end{figure}

Our $J$=2--1 map is the highest resolution CO map published so far of the Antennae representing 
the entire flux. We can identify both galactic nuclei, the IAR, and the western 
arc of emission in NGC\,4038 surrounding the galaxy from the northeast (weak emission) 
to the southwest (stronger emission). Furthermore, although buried in the map of 
total integrated 
CO\,(2--1) intensity, some of the {\bf S}uper {\bf G}iant {\bf M}olecular {\bf C}louds ({\bf SGMC}s) 
reported by Wilson et al. (2000) are indicated in the spectra (Fig.\,4) and channel maps (Fig.\,3) 
of the corresponding velocity ranges (for details see Sect. 3.2.). 
This also holds for the knot 9\arcsec northwest of 
SGMC\,1 in the map of Wilson et al. (their Fig.\,1), which is not addressed there but easily 
identified as a local maximum; we denote it as SGMC\,0 (see Fig.\,2). 

   \begin{figure*}
   \centering
   \resizebox{16cm}{!}{\rotatebox[origin=br]{0}{\includegraphics{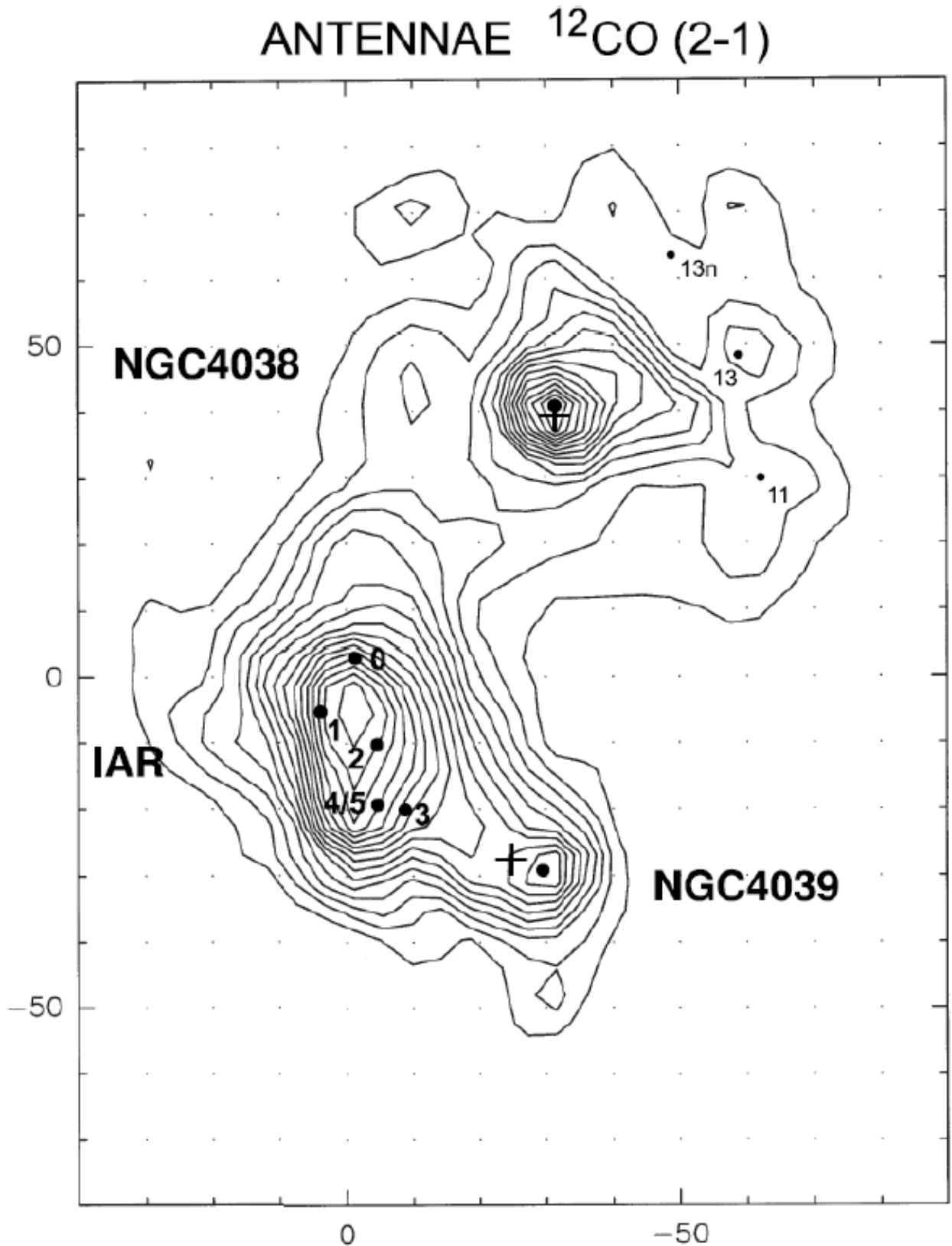}}}
   \caption{Map of integrated \TCO\,(2--1) line emission (vel. range 1300 to 
   1820 kms$^{-1}$ in contours of 20 to 170 K\,kms$^{-1}$ with 10 K\,kms$^{-1}$
   increment). Offsets in arcsec ((0,0)-position: $\alpha$(2000) = 12$^h$01$^m$55.2$^s$, 
   $\delta$(2000) = --18$^o$52$'$43$''$).  
   The crosses denote the galactic centres (2$\mu$m emission maxima of NGC\,4038 in the north and 
   NGC\,4039 in the south). Thick dots mark positions of the super giant molecular clouds (SGMCs) 
   of Wilson et al.   (2000), thin dots 
   the three strongest CO\,(1--0) maxima of Wilson et al. (2000) in the western arc of NGC\,4038 
   (notation of Hummel \& van der Hulst 1986).} 
   \label{ftwo}
\end{figure*}

$^{13}$CO\,(1--0) and (2--1) spectra were obtained at the three emission maxima (NGC\,4038, 
IAR and NGC\,4039). The pointing accuracy of these spectra ($\la$1$''$) was established 
by observing \TCO  in parallel. For these three emission maxima we  
present line ratios  (Table 1) which are needed to evaluate the physical state of the cold gas 
(see Sects. 5 and 6). 

\smallskip

The 870$\mu$m continuum map is displayed in Fig.\,5. The observed 870$\mu$m flux of NGC\,4038, 
the IAR, and NGC\,4039 is 195, 420, and 115 mJy, respectively. The total flux of the Antennae 
is about 730 mJy. The values of Zhu et al. (2003), being corrected for line emission, are 
slightly lower for NGC\,4038 and NGC\,4039 but only half of our value for the IAR. 


\bigskip

   \begin{figure*}
   \centering
   \resizebox{16cm}{!}{\rotatebox[origin=br]{-90}{\includegraphics{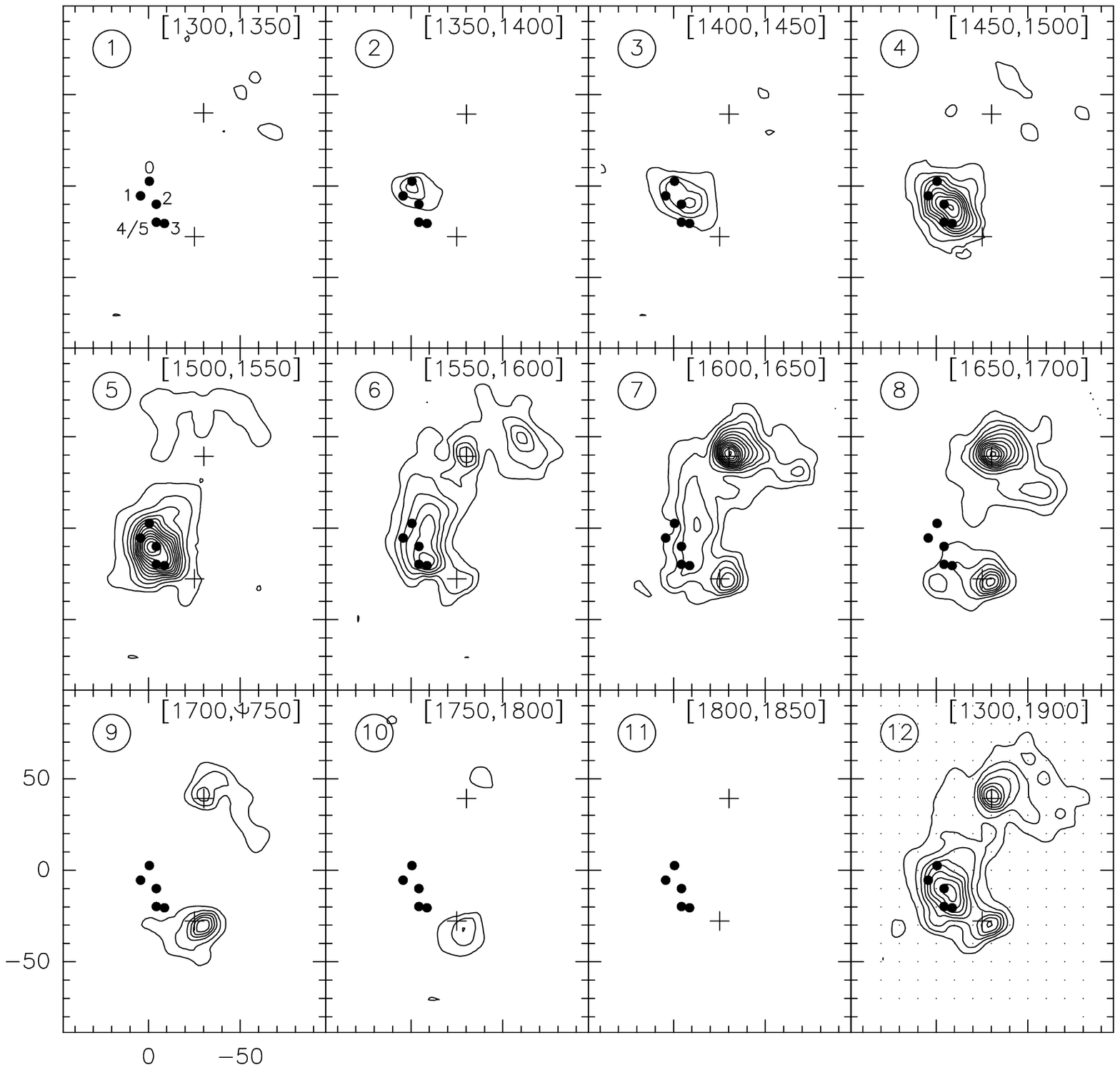}}}
   \caption{Channel maps of integrated $^{12}$CO\,(2--1) line emission. (0,0)-position 
   as in Fig. 2, offsets given in arcsec. 
   The velocity range of each map is indicated. Contours start at 6 
   K\,kms$^{-1}$ (3 K\,kms$^{-1}$ increment). The crosses denote the galactic nuclear positions
   (I-band centres), the dots indicate the SGMC positions (legend in Map 1).   }
   \label{fthree}
\end{figure*}

\subsection{Kinematics}

{\bf NGC\,4038}

The CO\,(2--1) spectra of Fig.\,4 show that the galactic disk gas of NGC\,4038 
is much less disturbed than that of the 
IAR and NGC\,4039. The shape of the central  CO\,(2--1) spectrum of NGC\,4038 deviates only 
slightly from a Gaussian. It does not allow us to distinguish different velocity components any 
more than does the series of channel maps (Fig. 3). The line shape (centred at 1640 km\,s$^{-1}$) 
and FWHM line width of 90 km\,s$^{-1}$  are in accordance with the nuclear region not being heavily 
disrupted by anomalous motions due to the encounter. 
The small velocity gradient in the disk from 1570 \kms in the southeast of 
NGC\,4038 to 1650 \kms in the west agrees with counterclockwise rotation (Hibbard et al. 
2001) and an almost face-on view.  The western arc shows a different velocity field rising 
from 1520 \kms in its northern part to 1600 \kms in the west and to 1700 \kms in the 
southwest. Some of the spectra are double-peaked and one can identify both the disk gas and 
the arc component; hence, the arc component should not be interpreted as a normal spiral 
arm-like feature but, more likely, as a streamer caused by the galaxy interaction.  

\smallskip

{\bf IAR}

In the IAR all the CO\,(2--1) spectra (Fig.\,4) have non-Gaussian shapes; some of them can easily be 
decomposed into two or more components. The CO\,(2--1) lineshapes look very similar to those of  
CO\,(1--0). The peak emission is centred at 
a lower velocity than that of both galactic nuclei (1520 km\,s$^{-1}$),  
but it ranges from 1350 km\,s$^{-1}$ to 
1720 km\,s$^{-1}$ in the wings. Lineshapes change drastically with position. 
The SGMCs  of Wilson et al. (2000) can be identified  - some of them only tentatively - in our channel 
maps of Fig. 3: SGMC\,1 (weakly) as well as SGMC\,2 and 3 in channel maps No.\,4 and 5, 
SGMC\,4 in map No.\,5, SGMC\,5 (weakly) in map No.\,6; SGMC\,0 north of SGMC\,1  
is indicated in our channel maps No.\,2 and 3, which may be associated 
with the strong component at 1390 kms$^{-1}$, most prominently seen 10$''$ north of the CO\,(2--1) 
maximum in the IAR (Fig. 4). In the southern part of the IAR we also begin to observe the disk gas 
of NGC\,4039, and IAR gas components show up throughout the entire eastern part of the NGC\,4039 
disk. 

Direct evidence of cloud collisions and infall are difficult 
to obtain since the 3-dimensional positions of the clouds are unknown. 

   \begin{figure*}
   \centering
   \resizebox{16cm}{!}{\rotatebox[origin=br]{0}{\includegraphics{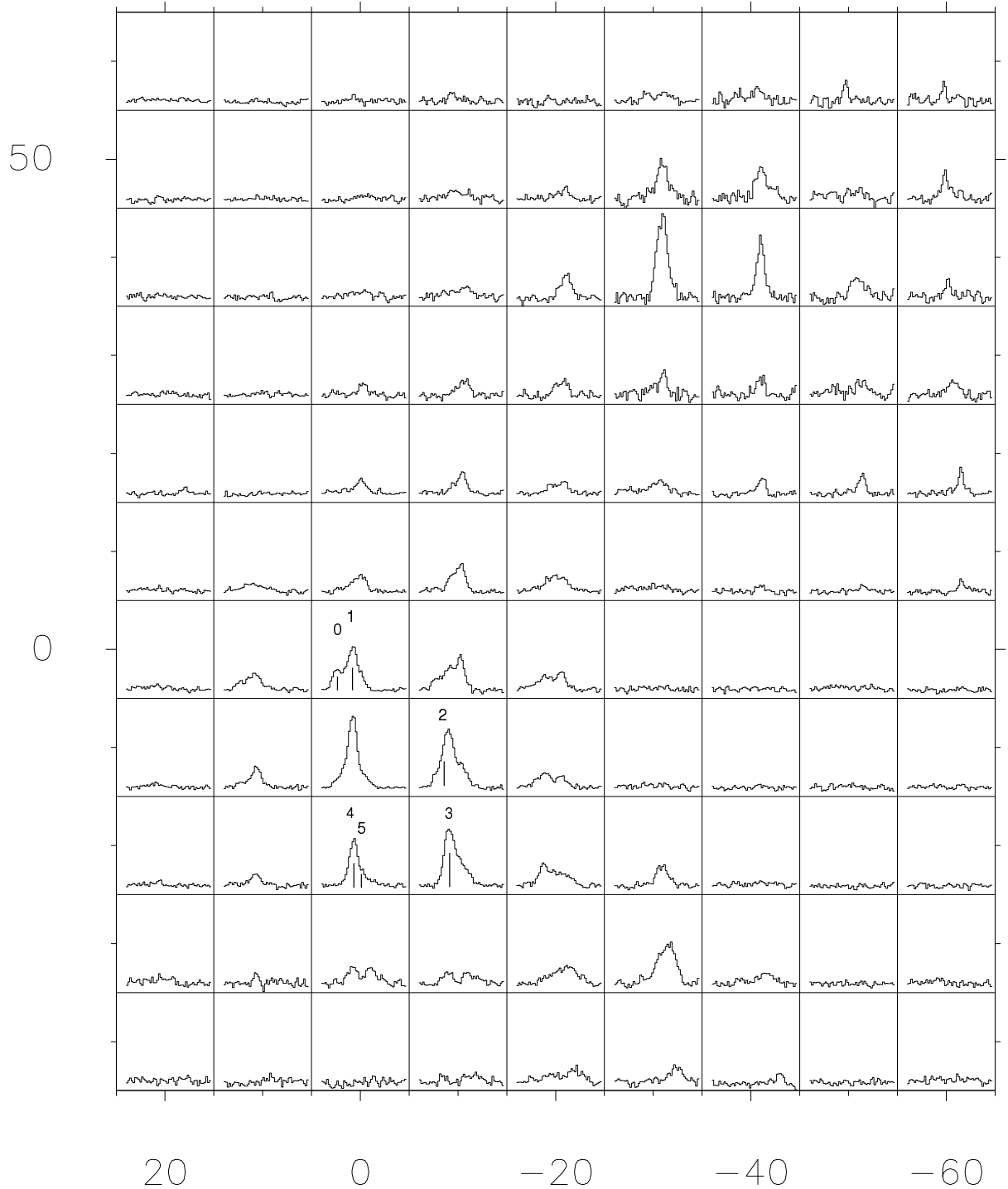}}}
   \caption{Map (central part) of \TCO\,(2--1) spectra. (0,0) position as in Fig.\,2, 
   offsets in R.A. and Dec. are given in arcsec. Velocity range 1200 to 1950 kms$^{-1}$; 
   the velocities of the SGMCs 1 to 5 after Wilson et al. (2000) and also the suspected 
   velocity of SGMC\,0 are marked; 
   intensity -- 0.15 to 1.5 K on a $T_{\rm mb}$ scale.}
   \label{ffour}
\end{figure*}

\smallskip

{\bf NGC\,4039}

The spectra attributed to NGC\,4039 (nucleus centred at 1665 km\,s$^{-1}$)
are mostly non-Gaussian and partly much broader than those 
of NGC\,4038. The observed velocity gradient across the disk from 1640 \kms (east) to 
1800 \kms (southwest) is larger than in NGC\,4038. Several spectra show more than one 
emission component; a bridge of CO connects NGC\,4039 to the IAR.
The disk of NGC\,4039 therefore appears to us more edge-on than that of 
NGC\,4038 and/or is more heavily disturbed by the interaction, the latter assumption being 
supported by the positional overlap with the IAR in the east. If part of the gradient 
is due 
to rotation, NGC\,4039 also rotates counterclockwise. At the southwestern edge we see some 
indication of a spiral arm. 

If NGC\,4039 is located in front of NGC\,4038 with respect to us (Hibbard et al. 2001), then 
both galaxies are approaching each other. This could enhance interaction processes in the 
future, possibly leading to a final coalescence. 


\section{Gas distribution and cloud morphology}

In order to discuss the physical state of the gas of the Antennae and its main 
heating mechanism(s) on the basis of our CO measurements, it is necessary (1) to compare our 
data with other CO observations and (2) to closely inspect the distribution of other tracers 
of the interstellar medium also thought to be related to star formation, 
at all available wavelengths from low-frequency radio emission to X-rays. To 
distinguish exact spatial coincidences and nearby associations and to provide 
a broader picture of the gas, dust, and star-formation activity in the Antennae,we attempt here 
for the first time to perform such an inspection to a very high positional accuracy of $\la$1$''$, 
employing maps with the highest angular resolution. 

We start with the CO maps, continuing with FIR and submm emission 
from the cold dust which should be correlated to the emission of cool molecular gas. Subsequently, 
we consider radio continuum, mid-infrared (MIR), near-infrared (NIR), and even the X-ray 
emission thought to be related to star formation.

\subsection{CO maps}

There is excellent agreement between our CO\,(2--1) map and the CO\,(1--0) interferometer 
map of Wilson et al. (2000), the CO map with the highest angular resolution available.
No positional difference of any feature could be observed between the maps. 
Because of this excellent agreement,
we will frequently use the interferometric map for morphological comparisons. 
We do not, however, consider their data to derive line ratios because 
(1) the resolution of the  Wilson et al. map is much higher than in our 
maps, making beam size corrections  uncertain, and because (2) Wilson et al. give 
only a rough overall estimate of their missing flux, which could vary from source 
to source. 

There is also good correlation to the single dish CO\,(1--0) map of Zhu et al. 
(2003), even including  their channel maps. However, we do not observe significant 
differences in the CO\,(1--0) and (3--2) line profiles at positions $\sim$20$''$ north of the IAR 
centre (see Sect. 6.1.1.). 
We also notice that the positions of SGMC 1 - 3 indicated in their channel maps 
(their Fig.\,7) are located at positions offset by up to 20$''$ from our 
map and that of Wilson et al. (2000), and SGMC 4/5 was left out. 


There is also good correspondence to the CO\,(1--0) 6$''$ interferometric map of 
Stanford et al. (1990) and the single-dish map of Gao et al. (2001), although the comparison 
with the latter map is limited by its low angular resolution (55$''$).

\subsection{FIR and submm emission}

At the IAR the distribution of cold dust emission (FIR and submm continuum) is 
in accordance with the CO maps within the positional uncertainties. Our 870\mic continuum map 
(Fig. 5) shows the same morphology as that of Zhu et al. (2003). Cold dust and molecular gas 
could be well-mixed here. 

   \begin{figure}
   \centering
   \resizebox{8.5cm}{!}{\rotatebox[origin=br]{-90}{\includegraphics{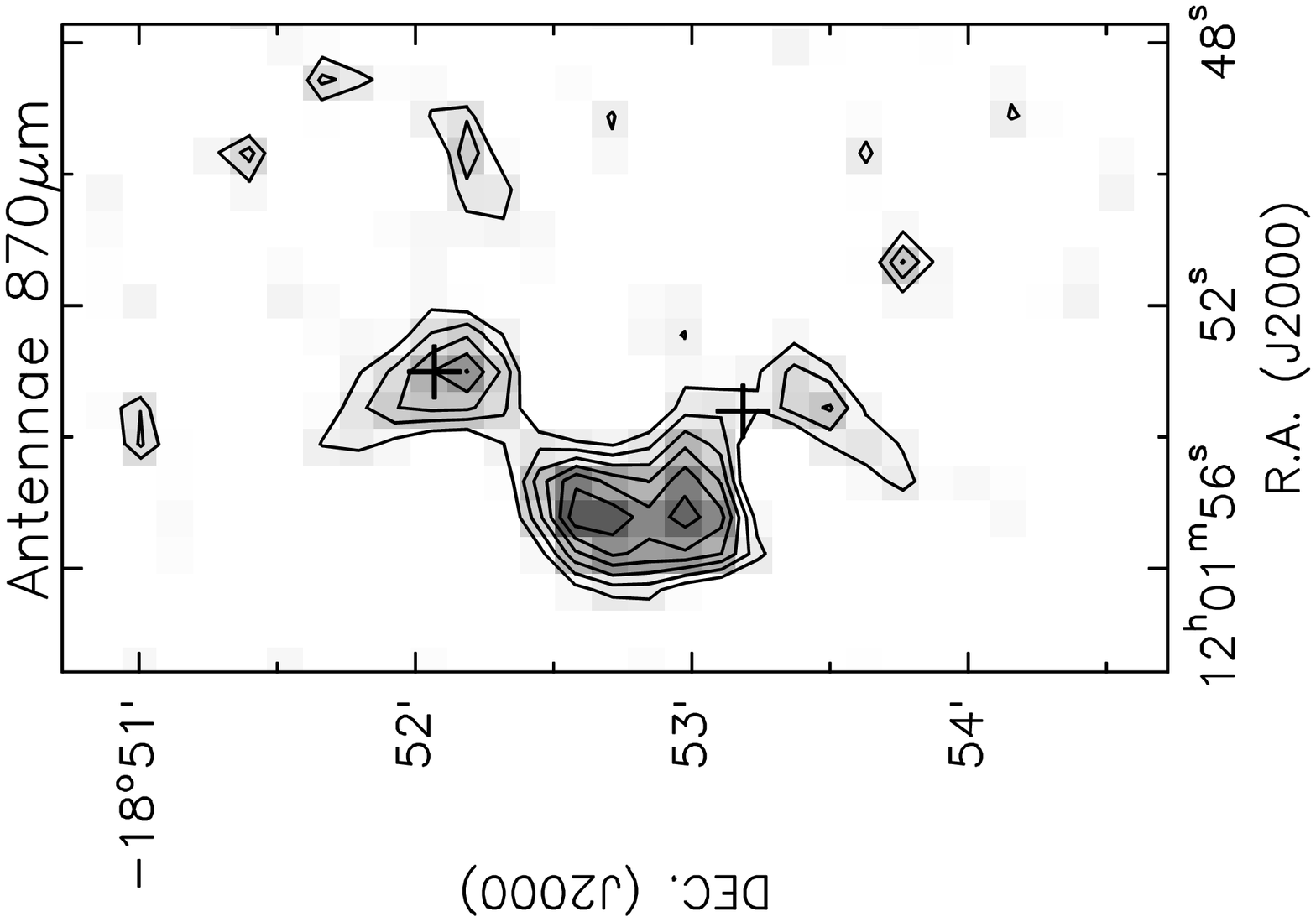}}}
   \caption{870\mic  map of the Antennae (22$''$ resolution, 60 to 160 mJy by 20 mJy, 
   1$\sigma$ = 20 mJy). Crosses denote the galactic nuclear positions.}
   \label{ffive}
\end{figure}

This correspondence is more difficult to observe for the other regions of the 
Antennae system. The emission peak of the NGC\,4038 nucleus  in our 870\mic map -  
in accordance with the 160\mic  emission by Bushouse et al. (1998) - is offset 
to the south compared to that of Zhu et al. (2003), but all three maps 
peak south of both the strong CO peak and the I-band centre. On the other hand, at 
60\mic the maximum is observed north of the CO and I-band peaks 
(Evans et al. 1997, Kuiper Airborne Observatory with a beam size of 17$''$), 
and the entire 60\mic  map of NGC\,4038 shows little correspondence to either 15\mic  
or 160/870\mic   emission. Not unexpectedly, the cold and hot dust trace different 
environments, which deserves further investigation. 


The cold dust in the NGC\,4038 western arc still remains an enigma: our 870\mic 
map does not have sufficient sensitivity; the area is not part of the map of Zhu et al. (2003), 
and 15\mic   and 60\mic   peaks do not coincide here at all. The 
nucleus of NGC\,4039 is very weak at 870\mic   but well-identified by tracers of warmer dust 
at 60\mic   and 15$\mu$m, typical of a disturbed galactic nuclear region.


\subsection{Radio continuum and MIR emission}

All the radio continuum maps with high angular resolution are morphologically identical 
to a precision of better than an arcsec (Hummel \& van der Hulst 1986 at 6/20 cm; 
Neff \& Ulvestad 2000 at 4/6 cm; Chyzy \& Beck 2004 at 6 cm). The radio 
emission peaks coincide  within $<$1$''$ with those of the MIR 15\mic emission (Mirabel 
et al. 1998), which is interpreted  to originate in warm dust. These emission 
maxima are associated with corresponding CO maxima, but none of the CO peaks 
is exactly identical (offsets 0.5$''$ -- 4$''$) to those of 15$\mu$m/radio emission 
(this is most obvious when  using the interferometric CO map of Wilson et al. 2000). 

The radio continuum emission 
has a thermal component from free-free radiation of H{\sc ii} gas and a nonthermal component 
that is likely to be tracing synchrotron radiation from supernova remnants (SNRs). Neff \& 
Ulvestad (2000) 
analysed the spectral indices ${\alpha}$ of the radio flux (S$_{\nu} \propto \nu^{\alpha}$). 
For the maxima of both galactic nuclei, they find indices
of about --0.6. For most of the sources in the IAR, they find flat spectra ($\alpha \ge$--0.5). 
They explicitly listed knots corresponding to SGMC 0,1,3, and 4/5 among the thermal sources. 
Rather flat spectra are also found for the knots of the western arc of NGC\,4038, except for the knot 
of area 11 of Neff \& Ulvestad ($\alpha$ = --0.77 $\pm$0.22, where, on the other hand,  Hummel \& 
van der Hulst (1986) get $\alpha$ = --0.4 $\pm$0.2). 
 

Flat radio spectra arise from H{\sc ii} regions with recent ($\le$10 Myr) star formation. 
Neff \& Ulvestad (2000) estimated $\sim$10$^4$ O\,5 stars needed to cause the thermal 
radio flux of the IAR. 
The exact coincidence of some of the radio knots with 
H$\alpha$ maxima (Amram et al. 1992; Whitmore et al. 1999;  a one-to-one correlation is not 
expected due to absorption of the optical light by dust) further supports this picture. 
The sources of the ionizing UV radiation must be 
clusters of early type stars that also heat the dust in their close neighbourhood; 
this produces the extremely good correlation to the 15\mic  map. The offset of the H{\sc ii} 
regions relative to the SGMCs implies that the SGMCs do not 
coincide with the starburst sites of the IAR and the western arc of NGC\,4038 produced by 
the galactic interaction, but they are sites of potential future star formation. Their 
neighbourhood to the clusters/superclusters causes their exposure to a strong UV radiation 
field that is discussed in more detail in Sect. 5.2.1. 

In the north eastern part of the Antennae (NGC\,4038 East), Hummel \& van der Hulst (1986) 
find a higher-than-average non-thermal fraction in the extended diffuse radio emission 
due to SNRs that had time to expand. This points to a higher age for this region compared 
to the IAR (see also Zhang et al. 2001). The high fraction of non-thermal emission  
near the two galactic nuclei indicates that SNRs are more numerous than in normal 
galactic centres. This is very likely a consequence of the galaxy interaction.

\subsection{Near-infrared}

NIR images of regions of enhanced star formation in the I-band are generally interpreted to 
be dominated by emission from 
red giant stars, so strong extended I-band emission should indicate older star birth 
sites in a starburst system and also older stars in 
the nuclear regions of both disk galaxies. On inspecting 
the I-band image of Whitmore et al. (1999) one realizes - not surprisingly - an offset 
of the galactic nuclei to the CO maxima of a few arcsec.  
The southwestern CO ``spiral arm'' of NGC\,4039 corresponds well to the I-band morphology. 
Considerably less I-band emission  from ``point sources'' is seen at the IAR. 
This should not be attributed entirely to high extinction because it also holds for regions 
where the extinction is low (Whitmore \& Schweizer 1995). 
Stronger I-band emission from  pointlike sources shows up (1) in NGC\,4038 East 
which has already been suggested as being older than the IAR starburst site (e.g. Zhang et 
al. 2001 and Sect. 4.3.), and (2) in 
the southern part of the NGC\,4038 western arc. For the latter region we observe  a 
general  correlation with CO  emission, but again {\it all} I-band maxima 
are offset to their corresponding CO cores (the offsets are random 
and cannot be artefacts caused by inaccurate pointings).

All this hints at a sequence of 
star formation from the most recent events in the IAR and perhaps the northern 
part of the NGC\,4038 western arc (where we find optical emission and CO but very little I-band 
emission) to the older sites of  activity in the southern part of the western arc and in 
NGC\,4038 East. This basically agrees with star cluster age distributions given by Kassin et al. 
(2003), Brandl et al. (2005), and Mengel et al. (2005). The locations of these sites of star formation 
all do not coincide with the CO clouds.

\subsection{X-Ray emission}

X-ray emission is thought to originate in SNRs or X-ray binaries (preferably the hard 
component), either from a hot diffuse interstellar medium or from shocked gas (mostly the soft 
component). When inspecting high angular resolution maps by Fabbiano et al. (1997, 2001), 
it is striking that X-ray emission (both pointlike sources and 
diffuse emission) is strong at both galactic nuclei, near 
NGC\,4038 East and in the southern part of the NGC\,4038 western arc but is fainter 
at the IAR. Its weakness at the IAR is likely to be real since it is observed for both kinds 
of sources and at positions of high, as well as low, extinction (for regions of low 
extinction see Whitmore \& Schweizer 1995). 

The diffuse X-ray emission in both galactic nuclei is stronger than in normal spirals. 
The hot-gas component should be close to pressure equilibrium with the cold gas component 
(Fabbiano et al. 2003). 
The softer spectrum of the NGC\,4038 nucleus is interpreted as originating predominantly in 
hot winds from star-forming regions; the NGC\,4039 nucleus also contains a harder spectral component 
attributed to X-ray binaries due to enhanced supernova activity (Zezas et al. 2002). 

Although we again find no exact positional coincidence with other images, the correspondence 
of X-ray bubbles (Fabbiano et al. 2001) with locations of suspected older starburst regions 
(see the I-band image of Whitmore et al. 1999) 
in NGC\,4038 East and the southern part of the western arc is noticeable. 
The correspondence of diffuse (non-thermal) radio and diffuse X-ray emission (Neff \& 
Ulvestad 2000) favours the idea that SNRs heat the hot coronal gas (Read et al. 1995).
On the other 
hand, the lower flux of X-ray emission in the IAR is consistent with (1) an absence of 
strong shocks on very large scales and (2) a low abundance or absence of evolved 
SNRs; if SNRs exist in the IAR, they must predominantly be fairly  
young ($\le$10$^7$ yr).

\section{Excitation analysis}

In this section we investigate the applicability of two different models to explain the 
excitation of the interstellar gas and to determine its physical parameters.

\subsection{LVG radiative transfer models}

Although frequently applied, a standard radiative transfer analysis using an LVG 
(Large-Velocity-Gradient) line escape-probability code to model observed CO 
transitions has certain disadvantages, 
particularly when used for large cloud complexes, also discussed by Schulz et 
al. (2001). LVG models use gas temperatures and 
densities as fixed input parameters despite the fact that for kpc-sized regions the 
temperature and density are unlikely to be constant. Temperature gradients 
considerably influence the excitation of molecular lines. 

Our own attempt to model our CO observations of the IAR yields no sensible results. The 
ratio of $^{13}$CO\,(2--1) and (1--0) can be fitted (both lines optically thin) for low 
gas temperatures and intermediate densities ($T_{\rm kin}$ = 30\,K: $n$ = 10$^3$ cm$^{-3}$; 
10\,K: $n$ = 5 10$^3$ cm$^{-3}$), 
but the ratio of $^{12}$CO\,(2--1) and (1--0) (see Fig. 6, lowest panel) demands a very low gas 
temperature and/or subthermal excitation for the CO\,(2--1) emission, which both contradicts 
the observed intensity of the CO\,(3--2) line. Because of the limited number of observed 
lines, the application of a two-component model makes little sense.  

Zhu et al. (2003) attempt to fit their CO observations of the Antennae. 
They report that, at least for the IAR, 
the observed line ratios require spatial temperature 
(and possibly also density) gradients. Even a two-component LVG 
model with different densities and temperatures 
yields, however, results that  are not well-confined. 
Moreover, their obtained CO line LVG models seem to suggest  - as in several other cases - 
that the densest gas component is warmer ($\geq$50K) than the less dense component, 
which should be questioned for kpc-size regions; 
in fact, investigations  of the central region of the Galaxy (e.g. H\"uttemeister et al. 1998) 
and of the galaxy IC\,342 (Schulz et al. 2001) yield the opposite: the dense gas component is 
cool. Furthermore, LVG models hint at neither the physical heating process nor the chemistry.

\subsection{Models of {\bf P}hoton {\bf D}ominated {\bf R}egions (PDRs)}

\subsubsection{Indications of PDRs}

By now there is continously rising evidence that soft UV radiation with 
energies below 13.6\,eV is  penetrating huge areas of both our own and 
external galaxies that is inhomogeneously filled with molecular clouds.  
Consequently, one should consider  applying PDR models 
in order to physically describe the interstellar matter. 
Goicoechea et al. (2004) analyse 
far infrared (FIR) spectroscopic observations of the Sgr B2 region of 
our own Galactic centre in terms of such models. Mochizuki \& Nakagawa (2000) 
successfully model observations of the [C{\sc ii}] and FIR emission of even larger areas 
of the inner Galactic plane. Wolfire et al. (1990) and Stacey et al. (1991), 
and, more recently, Malhotra et al. (2001) have analysed data of atomic fine 
structure lines and the FIR continuum in many galaxies in terms of PDR models. 
All these investigations show that PDR models are a useful tool for explaining 
observations of the diffuse interstellar matter not only on scales of individual 
clouds but also on the much larger linear scales characterising extragalactic studies. 

Models explaining the heating of dense molecular clouds with PDRs were established 
for galactic objects like the Orion nebula where atomic lines (e.g. [O{\sc i}], 
[C{\sc i}], [C{\sc ii}]) on the surface and molecular lines (e.g. CO) deeper inside the clouds 
are successfully modeled simultaneously (Stacey et al. 1993 and references therein). 
As indicated in the introduction, such analyses have also been successfully performed 
on kpc scales for the nuclear regions of the `normal` spiral galaxy IC\,342  (Schulz et al. 
2001), and the irregular `starburster`  M\,82 (Mao et al. 2000).

\smallskip

For the Antennae, many indications of the existence of widely-distributed 
PDRs have been found:

\begin{itemize}
\item
A map of 158$\mu$m [C{\sc ii}] emission (Nikola et al. 1998) 
reveals a distribution that is very similar to CO\,(1--0) (Gao at al. 
2001) observed with the same beam size of 55$''$. Nikola et al. show 
that the [C{\sc ii}] emission originates mostly in PDRs and that contributions from 
H{\sc ii} regions, as well as from extended low-density gas, are negligible (see also Kramer et al. 
2005). Nikola et al. obtain a value for the UV radiation field lower than 1000 $\chi_0$ 
(in units of the Draine-field $\chi_0$, see Sect. 5.2.2.), but, on the 
other hand, admit that their single-component fits for the galactic nuclei and for 
the IAR are very crude approximations, i.e. averages over huge areas. 
Directly comparing the [C{\sc ii}] and CO\,(1--0) (Gao et al. 2001) line intensities , both 
observed with a 55$''$ beam size, we obtain (in units of Jy)
a ratio of 2800 for the IAR, a value between those typical 
of normal non-starbursting galaxies (1000--1300) and those typical of starbursters ($\la$6000; 
see Stacey et al. 1993). For the nucleus of NGC\,4038 we find a value of 1800; for NGC\,4039, 
see below.

\item
Gilbert et al. (2000) analysed NIR spectra (2 to 2.5 $\mu$m) with 
a spectrometer slit crossing the mid-IR ISO peak of the IAR and the NGC\,4039
nucleus. The vibrationally excited H$_2$ lines extending over an area 
of 200 pc across the IAR are interpreted in terms of a PDR (rather than a shock) model 
with a UV radiation field of 5000 times the Draine field. The extent of the emission area suggests 
that the UV photons travel much farther than suggested by their free path length within 
a smooth medium of the density modeled ($\numd$ = 10$^4$ -- 10$^5$ cm$^{-3}$). This implies 
that the medium is extremely clumpy, a since well-established scenario for 
interstellar clouds (e.g. Wilson  \& Walmsley 1989). Br$\gamma$ emission indicates existing 
H{\sc ii} regions associated with the PDRs in the IAR, while this is weak or absent near the 
NGC\,4039 nucleus where star formation is currently weak. 

\item
Hot dust seen at 12 -- 17 $\mu$m (see Sect. 4.3.) is associated with the nuclear environment 
of the two galaxies, the western starforming ring of NGC\,4038 and the IAR (Mirabel et al. 
1998; Vigroux et al. 1996); it is heated by absorption of photons that are powerful 
enough to also ionize Ne, producing [Ne{\sc ii}] and 
[Ne{\sc iii}] lines that are observed to be well-correlated with the mid-IR emission.

\item
Fischer et al. (1996) performed FIR spectral imaging of fine structure lines 
of atomic oxygen and carbon. They find that the ratio of 
[C{\sc ii}] 158$\mu$m/[O{\sc i}] 63$\mu$m provides strong evidence that both lines arise 
from PDRs and derive a UV radiation field of up to 2500-times the Draine 
field (see Sect. 5.2.2.).

\item
Within the IAR, many super massive stellar clusters are observed as partly very young 
(Whitmore et al. 1999) and therefore as containing a large number of B stars providing the 
observed UV field that is heating the gas clouds mostly from the outside. 

\end{itemize}

By now, CO line radiation from a galaxy merger like the Antennae has never been analyzed 
with PDR models, but the variety of results cited above indicates that such an approach 
is reasonable.

\subsubsection{PDR model properties}

The ``KOSMA-$\tau$ PDR models'' applied here are described by K\"oster et al. (1994), St\"orzer et al. 
(1996), St\"orzer et al. (2000),  Zielinski et al. (2000) and R\"ollig et al. (2006). 
The most recent investigation slightly modified the PDR code after a comparison 
with equivalent codes of other research groups. 
The codes calculate model clouds illuminated by  UV photons between 6.2 
and 13.6 eV with a strength given in units of the local galactic interstellar 
radiation field  $\chi_0$ =  2 10$^{-4} \rm erg(cm^2 s\, sr)^{-1}$  (Draine 1978). 
%
An important improvement of PDR with respect to  LVG models is achieved by 
calculating the temperature gradients. 
This yields CO line ratios different from those of fixed-temperature models. 

The PDR models used here adopt plane-parallel layers of variable thickness illuminated 
by UV photons from both sides, as well as spherical clumps of variable radius illuminated  
from all sides. The plane parallel code (K\"oster et al. 1994) only permits 
a fixed density, whereas the spherical clump code (St\"orzer et al. 1996, R\"ollig et al. 2006) 
allows a fixed density and a 
power-law density distribution $n(r) = r^{-\beta}$ with $\beta$ as a free 
parameter; the clumps have a fixed size, the density starts with a value chosen at the surface 
and then rises with decreasing radius but is finally again fixed for the inner 10\% of the 
radius. Those models with different geometries are compared below. One reason to do 
this is that the results for IC\,342 (Schulz et al. 2001) were obtained assuming  plane-parallel 
and those for M\,82 (Mao et al. 2000) using spherical geometry. 

\smallskip
It is obvious that even the model with variable clump density operates with still very idealised 
conditions representing only first-order approximations to the real ones. For example, the 
structure of molecular clouds is rarely spherical but often filamentary, and the assumption 
of equal illumination from all sides is unrealistic. Furthermore, to model the few measured 
line ratios we are only able to  use an ``average'' PDR model. Averaging over our beam, 
filling several hundred parsec, it is impossible to obtain detailed physical properties for the 
several components of the interstellar gas that is observed (our smallest beam covers $\sim$1 kpc). 
In particular, it is difficult to model simultaneously the warm atomic gas (observed in atomic fine 
structure lines) and the cool molecular part of the clouds because their physical properties are 
very different. This is also discussed by Kramer et al. (2005), who  
apply PDR models to atomic fine structure lines and CO observations of the galaxies M\,83 
and M\,51. 

Even for the molecular part alone a difficulty 
arises when modeling extragalactic $^{12}$CO/\THCO  line ratios. 
These ratios are as low as 2 to 4 for the dense 
parts of galactic clouds observed with high spatial resolution, 
in agreement with model results. But for extragalactic 
objects the observed ratios always exceed 7. The line centre of \TCO   is optically 
thick for large portions of molecular clouds, whereas \THCO can only be  saturated 
in their densest parts. We therefore expect the low-density gas, which shows a pedestal 
of weak $^{12}$CO emission between the dense regions, to produce even weaker $^{13}$CO 
emission than the dense cores. 
Within our beam, we in fact observe regions often characterised 
not only by dense clumps but also by diffuse 
gas with comparatively high $^{12}$CO/\THCO line intensity ratios.  
As a consequence, observed ratios are on the order of 10 to 20 for  
the brightest parts of the Antennae and many other galaxies. This will probably also explain 
diffferences in $^{12}$CO and $^{13}$CO line maps observed in the future as seen in other galaxies, 
although both lines are emitted physically from the same regions. 

Nevertheless, this averaging problem has little 
influence on the resulting total hydrogen column density. 
Moreover, we can derive meaningful average parameters 
indicating the general physical state of the cloud complexes, in particular the temperature 
and density structure, which mainly determine the 
relative intensities of the different $J$-level transitions.

\begin{table*}
\label{ttwo}
\caption{{\bf Observed CO line ratios \& derived parameters}   }
\begin{tabular}{l c c c c c | c c r}
\hline
cloud & $R_{12}\frac{(2-1)^{1)}}{(1-0)}$ & $R_{13}\frac{(2-1)^{1)}}{(1-0)}$ & $R_{12}\frac{(3-2)^{1)}}
{(1-0)}$ & $R_{12/13}(1-0)^{2)}$ & $R_{12/13}(2-1)^{2)}$ &  \numd$^{3)}$  & $N_{\rm H_2}^{4)}$  & 
$M_{\rm H_2}^{5)}$ \\
\hline

   &    &  &  &  &  & 10$^4$ cm$^{-3}$ & 10$^{22}$cm$^{-2}$ & 10$^9$ \solmass \\
\hline
4038     & 0.8     & 0.7  & 0.66     & 11     & 10  & 3.5   & 7 (5)  & 1.8 \\
 $^{6)}$ & 1.05    &      & $<$0.9   &        & 15  &       &        &     \\
\hline
IAR      & 0.75    & 0.84 & 0.59     & 16     & 14  & 2.5   & 10 (6) & 3.  \\
 $^{6)}$ & 0.86    &      & 0.68     &        & 28  &       &        &     \\
\hline
4039     & 0.68    &       & 0.50     &        & 16  & 2.0   & 9 (4) & 1.3 \\
 $^{6)}$ & 0.85    &       & 0.80     &        & 17  &       &       &     \\
\hline
\end{tabular}
\\

$^{1)}$ ratio of (2--1) to (1--0) and (3--2) to (1--0) line peak intensities for \TCO and $^{13}$CO, 
resp. The (2--1) to (1--0) ratios are calculated for a 10.5$''$ beam size using the source sizes of 
Wilson et al. (2000). (3--2) to (1--0) ratios are taken directly from our data observed 
with a beam size of 22$''$ (errors are within 15\,\%) \\
$^{2)}$ ratio of \TCO \, and \THCO \, intensities as directly observed  \\
$^{3)}$ gas density of the cloud cores from the PDR models; for the different models applied see text 
of Sect. 5.2.3. Estimated accuracy $\leq$30\,\%.  \\
$^{4)}$ determined PDR model column densities (and, in brackets, values simply applying 
the standard $N_{\rm H_2}/I_{\rm CO}$ factor, see Sect. 6.1.2.) \\
$^{5)}$ calculated from determined column densities with cloud complex sizes by Wilson et al. (2000); 
inclusion of the gas of the extended CO pedestal emission raises the total gas mass to  
$\sim 10^{10}$ \solmass. 
Masses should be accurate to a factor $\leq$2, but note that the value might be more uncertain 
for NGC\,4039 (see Sect. 6.1.2.) \\
$^{6)}$ values calculated from the spectra of Zhu et al. (2003) for a 10.5$''$ beam size using source 
sizes of Wilson et al. (2000). \\
\end{table*}

\subsubsection{PDR model results}

{\bf CO lines}

Ratios of \TCO lines  emitted from different $J$-levels  and $^{12}$CO/\THCO line ratios are listed 
in Table 1. For two reasons we use observed peak main beam brightness temperatures \TMB   instead of 
integrated line intensities for our modeling:  
(1) most of the spectra are very complex showing several distinct components 
that are difficult to be accurately disentangled; (2) values of integrated 
intensity suffer considerably more from baseline 
problems than do \TMB peak  values. Integrated intensities are only taken to derive 
gas masses.

The average interstellar UV radiation field is confined to values 
between $\chi$ = 500$\chi_0$ and $\chi$ = 3000$\chi_0$ (see Sects. 5.2.1. and 5.2.2.). 
Higher values are only found in regions much smaller than 1 kpc (like H{\sc ii} regions), and 
lower values would contradict the reported IR and FIR line fluxes. The influence of $\chi$ 
onto the CO line ratios is discussed by St\"orzer et al. (2000). Raising $\chi$ without 
changing other parameters raises the $^{12}$CO/$^{13}$CO line ratios and lowers the CO\,(1--0) 
line with respect to the (2--1) and (3--2) emission. This trend can be compensated for by 
slightly varying other model parameters, but all these variations within a range confined 
by our observed CO line ratios leave the total H$_2$ column densities virtually unchanged. 
Nevertheless, our chosen value of $\chi$ = 1000 $\chi_0$ is strongly constrained by the 
observed ratio of [O{\sc i}] and [C{\sc ii}] emission (see below). 

In order to check the validity of the different subsequently evolved PDR models applied 
to different galaxies (see Sect. 5.2.2.), we 
compare the three different available ways of modeling PDRs, i.e. (1) plane-parallel, 
(2) spherical with fixed gas density, and (3) spherical with a power-law density distribution. 
For our specific sets of observed line ratios, we obtain specific 
results in accordance with St\"orzer at al. (1996). 
Small clump sizes (or thin layers for the plane-parallel cloud) are generally demanded by 
the observed $^{12}$CO/$^{13}$CO ratios. 
The models with a constant volume density (plane-parallel versus spherical) yield very 
similar values for clump diameters (resp. thicknesses of cloud layers) of about 0.1 pc, of gas 
column densities of 
about 5  10$^{21}$\cmsq  per velocity interval (1.2 kms$^{-1}$), and of inner CO core radii  
(where the density of the produced CO exceeds 50\% of its central value) of about 0.01 pc.

The spherical models operate with fixed total cloud sizes. 
Comparing  spherical models with constant or variable ($\beta$ = 1.5) 
gas volume density (see Fig. 6, panels a and b), 
the resulting total cloud size is 4 times larger in the latter case, but the resulting total column 
densities and the inner core radii (0.01 pc)  -  where the gas temperature is low and the CO density 
is above 50\% of its central value -  are all identical. Furthermore, all three PDR models yield 
within 20\,\% the same total gas density of the inner cores (about 3  10$^4$ cm$^{-3}$) as is 
given in Table 1. 

Regarding these agreements, on the one hand, and our 1-kpc-size beam and the small 
number of observed transitions, on the other, 
it appears unjustified to calculate models which are even more detailed, e.g. introducing a 
size spectrum of clumps (St\"orzer et al. 2000; Zielinski et al. 2000). 

   \begin{figure}  
   \centering
   \resizebox{8.5cm}{!}{\rotatebox[origin=br]{-90}{\includegraphics{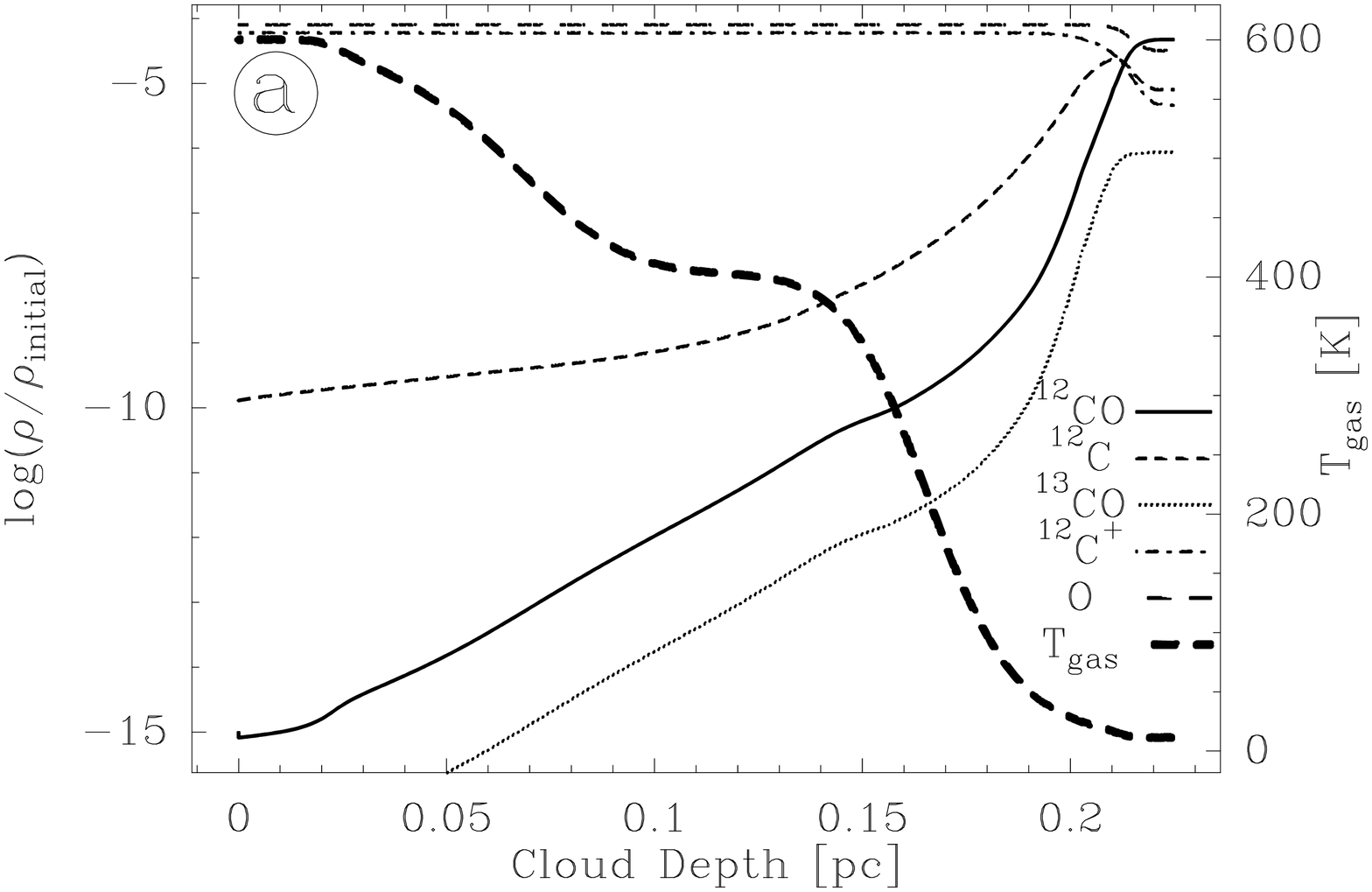}}}
   \resizebox{8.5cm}{!}{\rotatebox[origin=br]{-90}{\includegraphics{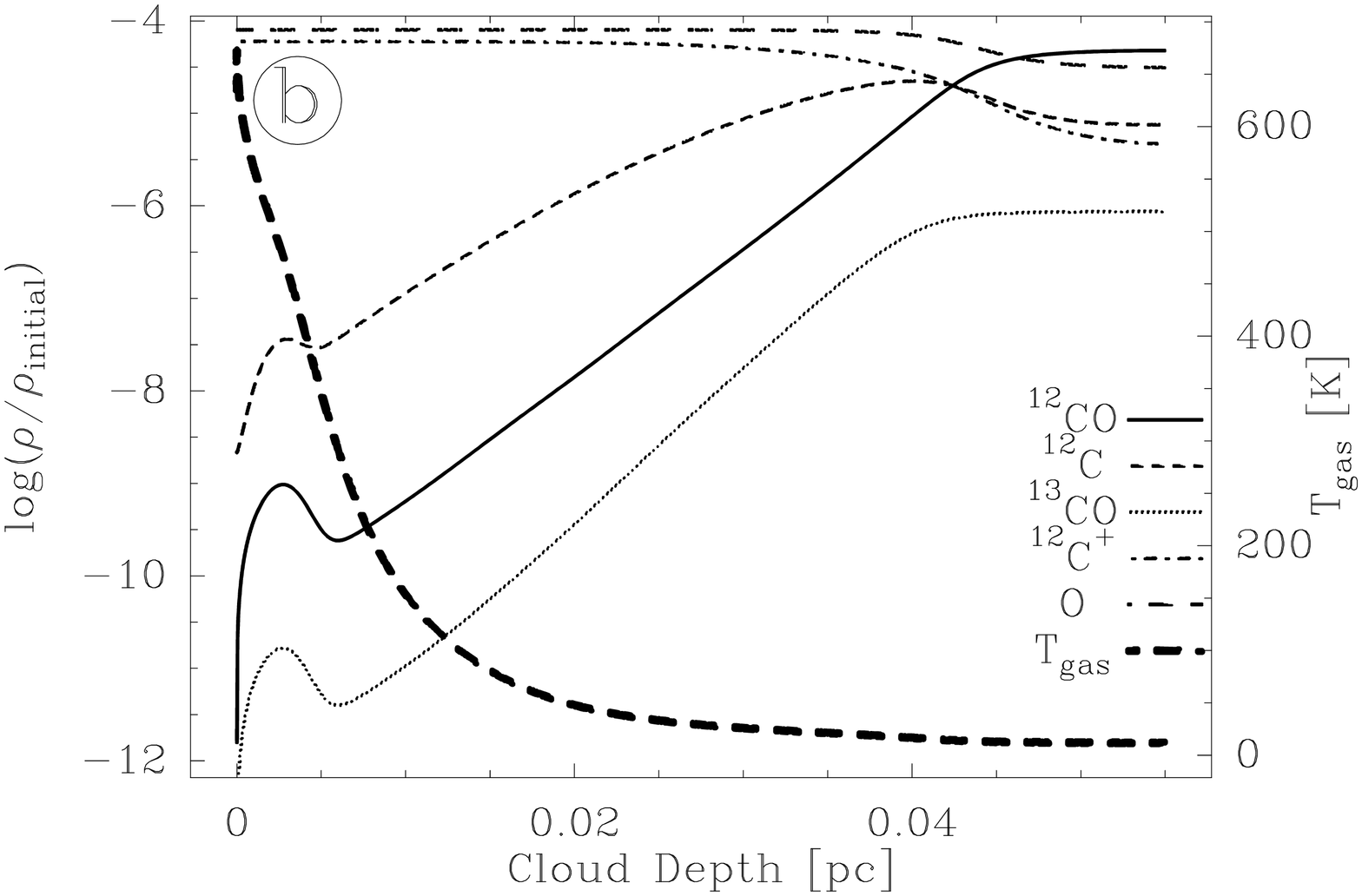}}}
   \resizebox{8.5cm}{!}{\rotatebox[origin=br]{-90}{\includegraphics{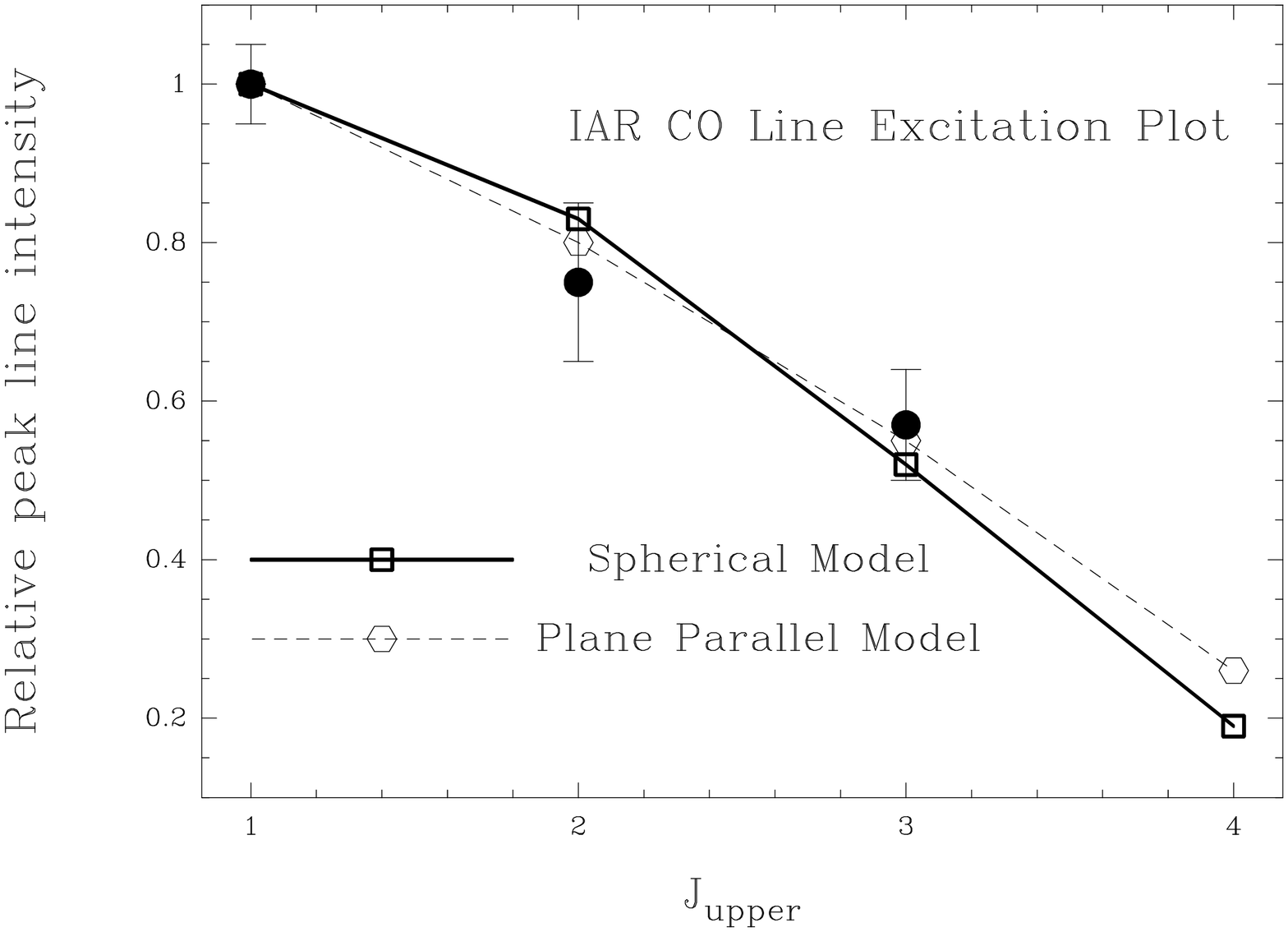}}}
   \caption{Modeled [C{\sc ii}], [C{\sc i}], [O{\sc i}], and CO (relative to H$_2$) densities and gas 
   temperatures versus clump depth (from surface to centre) 
    for both spherical PDR models of the IAR (a: variable total gas density from 1000 to 30000 cm$^{-3}$, 
    $\beta$ = 1.5;  b: constant total gas density of 3 10$^4$ cm$^{-3}$). $\chi$ = 1000 $\chi_0$. 
    The lowest panel shows the CO excitation plot. Open symbols: model; solid dots: observations.   }
   \label{fsix}
\end{figure}

For three areas where we have also observed \THCO spectra, we calculated specific model 
parameters including gas masses listed in Table 1. 
Gas densities contained in our models are rather well-constrained by the observed line ratios of the 
CO $J$-ladder. 
We note that, for the model with a variable density ($\beta$ = 1.5),  values at the outer edge of 
the model clump of $<$1000 \percc are unrealistic because they 
lead to densities for the small inner cloud cores being too low ($<$ 10$^4$ cm$^{-3}$) to 
yield observable CO line antenna temperatures for the obtained 
cloud radii (significantly confined by the $^{12}$CO/\THCO  line ratios). 
We also emphasise that, for sensible clump sizes (yielding moderate optical depths of the 
CO lines), a considerable fraction of gas within the inner cloud cores should not
considerably undercut the critical density to thermally 
excite the CO\,(3--2) line in the optically thin case (\numd = 5  10$^4$ cm$^{-3}$) 
since the CO\,(3--2) line peak intensity reaches 50--65\% of that of the (1--0) line. 
For the CO\,(4--3) line peak intensity we expect a value of 20\% to 30\% of the CO\,(1--0) value. 

\bigskip


{\bf Atomic lines}

The modeled line ratio of [O{\sc i}] 63$\mu$m and [C{\sc ii}] 158$\mu$m emission (1.3), 
which is very sensitive to the radiation field, agrees 
with the observed value of 1.4 (Fischer et al. 1996, their beam encompassing both galactic nuclei 
and  the IAR). Calculated line ratios of  [C{\sc ii}] and CO\,(1--0) line emission 
of the IAR and NGC\,4038 (3400 and 2300 in units of Jy, respectively) agree within 30\% with the 
observed values 
(2700 and 1800, respectively; Nikola et al. 1998 and Gao et al. 2001; for NGC\,4039 
such a comparison is not meaningful due to its position close to the IAR). Taking beam and possible 
source size differences into account, this agreement is very satisfactory, implying that all these 
lines should be emitted from the same regions. 


\bigskip

\section{Discussion}

\subsection{The physical state of the gas}

\subsubsection{Molecular excitation and $^{12}$CO/\THCO line ratios}


The good agreement between maps of different CO transitions (see Sect. 4.1.) also includes lineshapes. 
CO\,(1--0), (2--1), and  - for positions where the signal-to-noise ratio allows a decent 
comparison - also CO\,(3--2) profiles are quite similar, particularly for those positions where 
the lines show several components. 
Therefore, line ratios (and hence excitation conditions) should be similar for all these 
spatial components, including the IAR. In particular, this is also the case for the 
spectra 15$''$ to 25$''$ north of the IAR maximum where Zhu et al. (2003) report considerable 
differences for the CO\,(1--0 and (3--2) line shapes. They concluded that the 
velocity components exhibit completely different excitation conditions in this area. 

The gas must be highly inhomogeneous on {\it very} much smaller scales than our spatial resolution: 
model clumps have sizes of 0.1 to 0.5 pc. This is a typical clump size for galactic molecular clouds 
(in M\,17: 0.05 to 0.2 pc, Reid \& Wilson, 2006). 
Our models (see Sect. 5.2.3.) show that most of the CO emission arises from small moderately 
dense ($<$5  10$^4$ cm$^{-3}$) cores. Adding up the clumps at all observed velocities 
with total sizes of the cloud complexes as observed by Wilson et al. (2000) and correcting with 
filling factors derived from ratios of observed to modeled line temperatures, we end up with 
the masses given in Table 1. 

Most of the material in the interior of the clouds is at low temperatures ($<$30K). 
This accords with derived dust temperatures (Haas et al. 2000) and with the fact that no signs 
of an ongoing extreme starburst are observed in the CO clouds - the knots of intense thermal 
radio emission are {\it all} slightly offset from the molecular peaks (see Sect. 4.3.; 
this cannot be caused 
by pointing offsets since the offset directions are random).


From the diffuse synchrotron component of the radio emission, Hummel \& van der Hulst 
(1986) estimate a magnetic field strength of 40 $\mu$G, a high value for such an 
extended region very likely being generated by  compression of the ISM. 
We see evidence of a scenario where the generated young superclusters of stars 
meanwhile have partly disrupted their parent clouds and the 
ISM has been reorganised since then to its present distribution where starburst 
activity is rather low. 

The state of the gas close to both galactic nuclei is not very different from the gas of the 
IAR according to our 
analysis, although the star-forming rate compared to molecular gas mass should have 
been lower near the nuclei in the recent past than in the IAR. We derive this from inspecting the 
ratio of thermal radio continuum (Neff \& Ulvestad 2000) to CO flux.

\subsubsection{Molecular masses}

Our modeled gas masses (Table 1) agree with Wilson et al. (2000), adding up to about 
10$^{10}$ \solmass  for the entire system, which is about twice the value for our own Milky Way 
(Combes 1991). 
We should admit that the value for the NGC\,4039 nuclear region is an upper limit and quite uncertain, 
because our CO spectra at this position are possibly blended with emission from the IAR (it is 
not possible to separate IAR components from the broad spectra here).
Wilson et al. (2003) also investigated 
the cloud mass spectrum and obtained a range of SGMC masses of 2 10$^6$ to 10$^9$ \solmass    
for the IAR and find consistency of such very high cloud masses with the idea of the 
formation of a ``young super cluster'' population several Myr ago, as reported by Whitmore 
\& Schweizer (1995). Gao et al. (2001) obtain a total of 1.5  10$^{10}$ \solmass   using 
a standard  $N_{\rm H_2}/I_{\rm CO}$ conversion factor of 3 10$^{20}$ \cmsq(\Kkms)$^{-1}$. 
Also using this factor, 
Sanders \& Mirabel (1985) get 2.6  10$^9$ M$_{\odot}$ for the IAR, while Zhu et al. (2003) 
obtain $\sim$4 10$^9$ M$_{\odot}$ for the entire Antennae system. Stanford et al. (1990) get 
about 50\% of the values of Wilson 
et al. (2000) for both nuclei (for which they assume smaller sizes) and the IAR. 
If we  use the same standard conversion factor mentioned above to determine gas column 
densities derived  from integrated line intensities, we obtain the values listed in 
Table 1 in brackets. The values agree with the PDR modeling results within 
a factor of about 2 which we regard to be the anyhow accessible accuracy. 
Hence, we find no evidence of a significantly lower value than for Galactic disk clouds 
particularly for the IAR, in contrast to the results for the Galactic centre (Dahmen et al. 1998) 
and other galaxies that are supposed to be starburst galaxies such as M\,82 (Weiss et al. 2001) and 
NGC\,253 (Mauersberger et al. 1996) or even ultra luminous infrared galaxies (ULIRGs, see Solomon 
et al. 1997). The reason to meet such conditions within the IAR could be that 
the environment is different from that in a galactic nucleus. 
The gas close to the  centres of the Antennae galaxies, on the other hand, can originate to a 
large fraction from inflow of `normal` galactic disk material like what is encountered in the solar 
neighbourhood. However, this argument of gas inflow also holds for ULIRGs. A possible 
explanation for this difference between galaxies undergoing strong nuclear starbursts and 
the central regions of NGC\,4038 and NGC\,4039 may be the linear scale: The low 
$N_{\rm H_2}/I_{\rm CO}$ ratios in ULIRGs and the Galaxy refer to the central few 100 pc, 
while the regions studied here are larger. Consequently, 
if regions exist close to the centres of the Antenna galaxies with conversion factors 
such as found in NGC\,253 or in ULIRGs, they must be considerably smaller than 1 kpc (the 
size covered by our smallest beam). 

The total mass of the ionized gas in the radio knots (thermal component) is about 10$^8$ \solmass 
(Hummel \& van der Hulst 1986). The total dust mass is 10$^8$ \solmass  (Haas et al. 2000) in 
agreement with a normal galactic gas-to-dust ratio of order 100. 

Zhu et al. (2003) modeled their CO observations using an LVG model. 
For the inner kpc of NGC\,4038 they find 
\TKIN = 43\,K and \numd = 3.5  10$^3$ cm$^{-3}$. Compared to our PDR model results this density 
appears to be too low, also in the light of both their reported CO\,(3--2)/(1--0) 
line ratio of 0.9 and the critical density of the CO\,(3--2) line. 
For the IAR, Zhu et al. suggest a two-component solution with (A) 
\TKIN  = 30 -- 130 K and \numd = 1 -- 8  10$^3$ cm$^{-3}$, and (B) 
\TKIN  = 40 -- 200 K and \numd  $\ge$3  10$^4$ cm$^{-3}$. Applying a CO/H$_2$ abundance ratio 
of 10$^{-4}$ they obtain from their LVG modeling  molecular gas masses of the IAR 
of about 1.5 10$^8$ M$_{\odot}$, 
which is a factor of 10 lower than what they obtained with the standard $N_{\rm H_2}/I_{\rm CO}$ 
factor (1.8 10$^9$ M$_{\odot}$). The mass from their LVG modelling is also much lower than the 
values of the other studies. 
Zhu et al.  conclude that CO could be underabundant by a factor of $\sim$10, 
which seems not to be very likely in view of the star formation activity in the IAR. 
Furthermore, the result that the denser component (B) is warmer than component (A) 
is questionable when averaging over huge areas of kpc size (see Sect. 5.1.), also in view of the 
derived low dust temperatures (NGC\,4038 and IAR: 30K, NGC\,4039: 21K; see 
Amram et al. 1992, Haas et al. 2000).

If our derived gas masses (Table 1) are correct, huge amounts of gas are located within a radius 
of about 1 kpc of the galactic nuclei, almost a factor of 100 more than in the same area of 
the Milky Way. This agrees with the results of the Casasola et al. (2004) study of 
the gas content of 1038 interacting galaxies. Although no direct 
observational evidence exists in the case of the Antennae, it is plausible that the  
strong disturbance of the kinematical balance of a galaxy is resulting in mass inflow 
toward the central region.

\subsection{Alternative heating processes of the gas}

We have identified the soft UV radiation from B stars as one of the major heating sources for the 
gas of the Antennae that causes widespread PDRs (see Sect. 5.). In this section 
we try to investigate the contributions of other possible heating mechanisms, particularly 
for the IAR. 

\bigskip

{\bf Shocks}

Indicators of the presence of shocks in the ISM may be  vibrationally-excited H$_2$ 
emission,  X-ray emission, and  emission from molecular species produced in shocks like 
SiO. Indirect evidence for shocks can be provided by the existence of numerous SNRs. 

Gilbert et al. (2000) have analyzed near-IR spectra (2.03 - 2.45 $\mu$m   
containing several H$_2$ lines at very different excitation levels) 
observed in a compact star cluster of the IAR (at the position 
of the brightest 15$\mu$m knot identified by Mirabel et al. 1998, $\sim$20$''$ east of the 
NGC\,4039 nucleus) entirely in terms of a clumpy PDR. This is strongly supported by low values of 
[S{\sc ii}]/H${\alpha}$ line intensity ratios observed in the optical knots of the IAR, which are 
indicative of photoionization of the H{\sc ii} gas rather than shock heating (Whitmore et al. 2005). 
Gilbert et al. (2000) take shocks by SNRs or merger-induced cloud 
collisions into account only for the NGC\,4039 nucleus 
because they find that, due to the absence of Br$\,{\gamma}$ emission, 
star formation activity should currently be low. 


Analysing the radio emission of the IAR, 
Neff \& Ulvestad (2000) have shown that most SNRs must be very young: 
They estimate that there are $\sim$10$^4$ O\,5 stars needed to produce the thermal component. 
Using life times of a few Myr for these stars one would expect a SN rate of $\sim$ 0.01 per 
year if they were formed continually over the past few Myr. On the other hand, the 
nonthermal radio component implies a more than 20-times higher SN rate. The discrepancy 
can only be solved by assuming that all the O stars have been formed within a short period of time
a few Myr ago, thus implying that their SN shocks have not yet propagated over large distances 
to heat up a mayor fraction of the molecular clouds. 


Furthermore, the X-ray emission is not very strong throughout the IAR, 
particularly in the region near the CO line maxima and the locations of the youngest star 
clusters (Zhang et al. 2001); it is  stronger in other regions thought to host older sites 
of star formation. 
It might be expected in a merger system that on small scales the hot gas is not everywhere 
in pressure equilibrium with the cooler matter and  shocks might be present in a system 
with a recent starburst. But all the observed phenomena lead us to the conclusion that at present 
massive shock heating  either due to SNRs or due to cloud collisions is not likely to be the 
{\it dominant} heating source for the SGMCs of the IAR. Nowhere does its dense gas show 
signs of strong shocks on large scales. Nevertheless, it cannot be excluded that  heating by 
shocks plays a more important role in other parts of the merger system.

\bigskip

\bigskip

{\bf Cosmic rays (CRs)}

The dominant source of CR protons contributing to the heating of molecular clouds is 
assumed to be SNRs. 
Evidence of large numbers of extended 
(and therefore older) SNRs detected by hard X-ray emission is more rarely 
found in the IAR than in regions that are thought to be older: 
NGC\,4038 East and the southern part of the NGC\,4038 western 
arc (see Read et al. 1995). 
The diffuse soft X-ray thermal emission ($\sim$50\% of the total X-ray emission) thought 
to be generated by stellar winds and SN explosions heating the thin hot component of the 
ISM suggests a SN rate of 0.006 yr$^{-1}$ for the Antennae system (Fabbiano et al. 1997). 
The compact SNRs postulated by Neff \& Ulvestad (2000) from non-thermal radio emission (see 
above) are still too young to transport their particles over large distances through the IAR. 

Farquhar et al. (1994) correlated gas 
temperature and abundance of several atomic and molecular species versus the CR rate. Taking 
derived dust temperatures representative of the inner parts of molecular clouds of 
$\leq$30K would yield a maximum CR flux of $\leq$150-times that of the galactic value if 
the clouds were predominantly heated by CRs. Taking this estimate for the CR flux, the ratio of 
abundances of O{\sc i} and C{\sc ii} should be higher than 10$^3$ (Farquhar et al.). This 
conflicts with measured [O{\sc i}] and [C{\sc ii}] lines (Fischer et al. 1996), 
which are otherwise found to be typical of PDRs. 

A good correlation between CO and radio continuum emission is difficult to  interpret in terms 
of dominant CR heating of the molecular clouds even if an exact distinction of thermal (due to 
hot young stars) and nonthermal (due to SNRs) radio emission is achieved, because 
one also tends to find large quantities of O and B stars at sites with supernovae. 

Other ``caveats'' come from the following problems. 
(1) Charged CR particles are affected by magnetic fields (and also 
MHD waves) and therefore could fail to effectively penetrate cloud complexes. (2) Models of CR 
heating (Suchkov et al. 1993, Farquhar et al. 1994) predict that the molecular gas temperature is 
almost independent of gas density, although temperature gradients as a function of density are 
widely observed in molecular clouds, a situation which has led 
many investigations to invoke at least two gas components whenever their 
modeling was unable to implement temperature (and density) gradients. 

The argument that the time scale of CR diffusion, particularly in the presence of magnetic fields, 
is shorter than the time scale for producing UV radiation from late B stars might not be 
important since all activity is very young in the case of the Antennae. 
Nevertheless, the contribution of CRs to the heating of the SGMCs in the Antennae is 
difficult to determine; it is likely that it is not dominant in the IAR, but 
it may be different locally in other parts of the Antennae.

\bigskip

{\bf Ambipolar diffusion} 

Heating the extremely clumpy molecular cloud gas by ambipolar diffusion (ion slip) can not 
be completely excluded anywhere, but clear evidence in favor or against this mechanism 
is difficult to obtain. When a cloud is compressed, the magnetic field strength B 
will increase with gas density by n$^y$ with {\it y} $\leq$ 2/3 . As the friction of ionized and 
neutral gas also increases, {\it y} will be reduced, which decreases the heating efficiency. 
If, furthermore, the magnetic field gradient is smallon scales compared to the size of a molecular 
cloud, heating by ion slip weakens further (Scalo 1977). Hence, ambipolar diffusion might 
be quite important in the last stages of cloud-collapse forming stars, but it should not dominate 
the giant molecular cloud complexes discussed here.

\section{Conclusions}

%
%
From our CO line analysis probing the molecular gas  of the Antennae system, we draw the 
following conclusions in conjunction with many other data in the literature: 

{\bf (1)} We observe CO maxima embedded within a pedestal of CO emission at the positions 
of the super giant molecular cloud complexes (SGMCs, Wilson et al. 2000) that are {\it much} 
larger than Sgr B2 near our Galactic centre. 

{\bf (2)} Most of the IAR gas is found at significantly lower velocity than the gas near the 
two galactic nuclei, indicating that it is expelled from the merger system. One can find several 
velocity components that can be identified with the SGMCs. 
The NGC\,4038 disk shows a smooth velocity gradient, whereas the large western arc is kinematically 
separated from the disk and should be interpreted not as a spiral arm but as a separated filament 
caused by the interaction. The disk of NGC\,4039 shows the same east-west gradient but a large 
fraction of its velocity structure is contaminated by IAR gas components, and its nuclear spectrum 
looks much more disturbed by the encounter than does the NGC\,4038 spectrum. 

{\bf (3)} Classical large-velocity-gradient (LVG) models do not hint at 
the physical mechanism heating the gas. Recent model approaches yield poor fits and 
parameters for the molecular gas of the Antennae which are not very conclusive. Also, it 
results in a total gas mass that is 10-times lower than in several other studies.

On the other hand, in the literature we find widespread evidence of PDRs in the 
Antennae system. Our PDR model is able to reproduce the observed CO line ratios with a single component. 
All three of our model geometries  yield consistent results constraining the gas parameters within 
sensible limits. The model clouds are {\it much} smaller than 
the SGMCs, which indicates that the gas is highly clumped, thus allowing the UV radiation to 
penetrate large areas. Observed and modeled CO and [C{\sc ii}] line intensities are in 
accordance with the idea that a large fraction of the [C{\sc ii}] emission  originates in 
PDRs forming the outer layers of the identified molecular cloud complexes that in some 
cases are also associated with regions of moderate thermal free-free emission, i.e. H{\sc ii} 
regions. The CO excitation conditions averaged over a kpc scale appear to be approximately uniform 
across the merging system. We predict a CO\,(4--3)/(1--0) line intensity ratio of 0.18 to 0.26. 

Regarding the successful application of PDR models for the Antennae system as well as for 
galaxies like IC\,342 (Schulz et al. 2001) and M\,82 (Mao et al. 2000), 
one might suspect that PDR scenarios are generally found to be widespread in the interstellar 
medium of spiral galaxies. 

{\bf (4)} The modeled cloud cores have a moderately high density up to 4 10$^4$ cm$^{-3}$ 
at low temperature ($\leq$25K). The clouds show no signs of intense starburst activity: 
like thermal radio or MIR emission, such signs are 
{\it all} observed to be offset from the molecular peaks. The molecular gas mass contained in the 
nuclear 1 kpc of NGC\,4038 and NGC\,4039 exceeds that of the central 1 kpc of the 
Milky Way by a factor of almost 100, and the gas, including the IAR, does not seem 
to deviate much from a normal Galactic disk gas $N_{\rm H_2}/I_{\rm CO}$ factor. 
The gas/dust ratio appears to be the normal Galactic value of $\sim$100. The bulk of the gas of the 
SGMCs found in the IAR adds up to $\sim$3 10$^9$ $\rm M_{\odot}$,  the total molecular gas mass 
of the system ($\sim$10$^{10}$ $\rm M_{\odot}$) is about twice the total gas mass of the Milky Way.

{\bf (5)} We tried to examine the contribution of non-PDR heating sources for the gas 
seen in CO. 
Cosmic rays are likely to contribute to the heating of the cold molecular gas, 
but at least within the IAR they do not appear to be the 
dominant heating source because of the low X-ray flux here. In addition, 
the ratio of the [O{\sc i}] and [C{\sc ii}] line intensities does not support a high CR flux. 
Turbulent friction and shocks can also not be ruled out, although 
at least within the IAR there is no strong evidence for a dominant contribution 
if the interpretation of H$_2$ 
line emission in terms of PDRs is correct (Gilbert et al. 2000). Moreover, 
a gas temperature of $\sim$25K for the interior of the clouds obtained from our model 
does not favour the existence of strong extended shocks.

{\bf (6)} Careful comparison of maps for various indicators for star formation points to various 
ages for the different starburst areas in the Antennae system. In order to obtain a more 
detailed picture of the evolution of this prototypical system, it appears worthwhile to use 
probes of different states of excitation at much higher spatial resolution. In particular, 
very accurate intensities of molecular line series observed with interferometers are 
needed to obtain sensible line ratios to investigate the conditions on scales considerably 
smaller than 1 kpc.


\acknowledgements
We would like to thank the HHT staff members 
for their engaged support of our sometimes very 
experimental  project.
Ute Lisenfeld heped with the observations at the IRAM MRT. 
Frank Bertoldi, Carsten Kramer, Karl Menten, J\"urgen Stutzki, and Axel Weiss, 
as well as the anonymous referee, contributed useful discussions and comments. 

{}

\end{document}